\documentclass[12pt]{article}

\usepackage{graphicx}
\usepackage{hyperref} 
\usepackage{verbatim} 
\usepackage{url}
\usepackage{epstopdf}
\usepackage[normalem]{ulem}
\usepackage{color}
\usepackage{ulem}
\usepackage{dsfont}
\usepackage[utf8]{inputenc}
\DeclareUnicodeCharacter{2010}{-}
\usepackage{float}
\usepackage{amsmath}
\usepackage{comment}
\usepackage{breqn}
\usepackage{amssymb}
\usepackage{bbold}
\usepackage{times}
\renewcommand\vec[1]{\ensuremath\boldsymbol{#1}}
\newenvironment{sciabstract}{%
\begin{quote} \bf}
{\end{quote}}

\newcounter{lastnote}

\title{Generation of spin currents by a temperature gradient in a two-terminal device}

\author
{Rafael E. Barfknecht,$^{1,2\ast}$ Angela Foerster,$^{3}$ Nikolaj T. Zinner$^{4,5}$ and Artem G. Volosniev$^{6}$\\
\\
\normalsize{$^{1}$INO-CNR Istituto Nazionale di Ottica del CNR, 50019 Sesto Fiorentino, Italy}\\
\normalsize{$^{2}$LENS, European Laboratory for Non-Linear Spectroscopy, 50019 Sesto Fiorentino, Italy}\\
\normalsize{$^{3}$Instituto de F{\'i}sica da UFRGS, Av. Bento Gon{\c c}alves 9500, Porto Alegre, RS, Brazil}\\
\normalsize{$^{4}$Department of Physics and Astronomy, Aarhus University, Ny Munkegade 120, Denmark}\\
\normalsize{$^{5}$Aarhus Institute of Advanced Studies, Aarhus University, DK-8000 Aarhus C, Denmark}\\
\normalsize{$^{6}$Institute of Science and Technology Austria, Am Campus 1, 3400 Klosterneuburg, Austria}\\
\normalsize{$^\ast$To whom correspondence should be addressed: barfknecht@lens.unifi.it}
}

\date{\today}

\begin{document}
\baselineskip24pt

\maketitle

\section*{Abstract}
\begin{sciabstract}
Theoretical and experimental studies of the interaction between spins and temperature are vital for the development of spin caloritronics, as they dictate the design of future devices. 
In this work, we propose a two-terminal cold-atom simulator to study that interaction. The proposed quantum simulator consists of strongly interacting atoms that occupy two temperature reservoirs connected by a one-dimensional link.  
First, we argue that the dynamics in the link can be described using an inhomogeneous Heisenberg spin chain whose couplings are defined by the local temperature. Second, we show the existence of a spin current in a system with a temperature difference by studying the dynamics that follows the spin-flip of an atom in the link. A temperature gradient accelerates the impurity in one direction more than in the other, leading to an overall spin current similar to the spin Seebeck effect.
\end{sciabstract}

\section*{Introduction}

The coupling between charge and heat currents -- the thermoelectric effect -- was discovered more than two centuries ago. Today, this effect is a standard topic in physics textbooks~\cite{Goldsmid2010}, and is at the heart of thermoelectric generators and thermocouples. The spin thermoelectric effect -- the interaction between heat and spin currents -- has a much shorter history~\cite{bauer2012,Boona2014,YU2017}, but it already demonstrates the potential to complement the success of its older sibling. Spin thermoelectrics encompasses spin Seebeck~\cite{Uchida2008}, spin-dependent Seebeck~\cite{Johnson1987}, spin-dependent Peltier~\cite{flipse2012}, and related physical phenomena, which may lead to conceptually new devices based on the spin degree of freedom. While solid-state setups have provided crucial insight into the problem of spin transport, their limited degree of tunability does not allow one to go beyond the parameter regime given by the material at hand. Therefore, a logical next step is to explore spin thermoelectric effects using quantum simulators, in particular, cold-atom simulators, which provide a highly controllable environment for studying transport phenomena~\cite{bloch1,Chien2015}. Features of cold-atom systems such as the possibility to realize low-dimensional geometries and to control interactions are particularly favorable for the study of spin transport. 
 
It has been proposed to simulate certain features of spin caloritronics using three-dimensional cold gases where spin-up particles are separated from spin-down particles by applying a spin-dependent temperature gradient~\cite{PhysRevA.85.063613,Wong2012}. 
However, those approaches are experimentally challenging, especially in a strongly interacting regime. In this work, we propose a two-terminal device to study spin thermoelectrics for strongly interacting systems. Two-terminal cold-atom systems are a state-of-the-art interpretable platform for studying transport phenomena~\cite{Krinner2015,GRENIER2016,Krinner_2017,Lebrat2019}, and a playground for developing sophisticated quantum technologies. 
To illustrate our idea, we study the dynamics of a strongly interacting two-component atomic system in a small one-dimensional link between two reservoirs at different temperatures. The system has a large population imbalance, and we focus on the dynamics of the minority component.  To describe the system, we employ a basic theoretical model, which does not manifest any heat and charge transfer (in contrast to the DMRG study performed in \cite{PhysRevB.95.115148}, for instance). Instead, it exhibits certain features of the magnon-driven spin Seebeck effect~\cite{Adachi_2013}. In particular, it contains microscopic physics of a spin current in a ferromagnetic insulator. 

In our study, we rely on a correspondence between strongly interacting one-dimensional systems and spin chains. For zero temperature, this correspondence was studied theoretically in Refs.~\cite{deuretz1,deuretz2,artem1,artem2,pu1,massignan,xiaoling,xiaoling2,barfk3, Barfknecht2018}, and experimentally in a few-body setup of Ref.~\cite{jochim3}. In the present work, we use it to study the dynamics of a finite-temperature system. In most cases, we will focus on the dynamics that follows a single spin flip in the link, namely the ``impurity" model, which connects our findings to the physics of spin excitations in cold-atom simulators~\cite{fukuhara}. However, our observations can be generalized easily to a few spin impurities placed in the reservoirs. 

For convenience of the reader, let us summarize the main findings of the paper: a) we introduce an effective model for studying the time dynamics of a strongly interacting one-dimensional system at finite temperatures. This allows us to circumvent the need to explicitly include excited states in the analysis; b) Using this model, we present a microscopic theory in which a spin current is induced by a temperature gradient. 

\section*{Results}\label{results}

\begin{figure}
\centering
\includegraphics[width=0.6\textwidth]{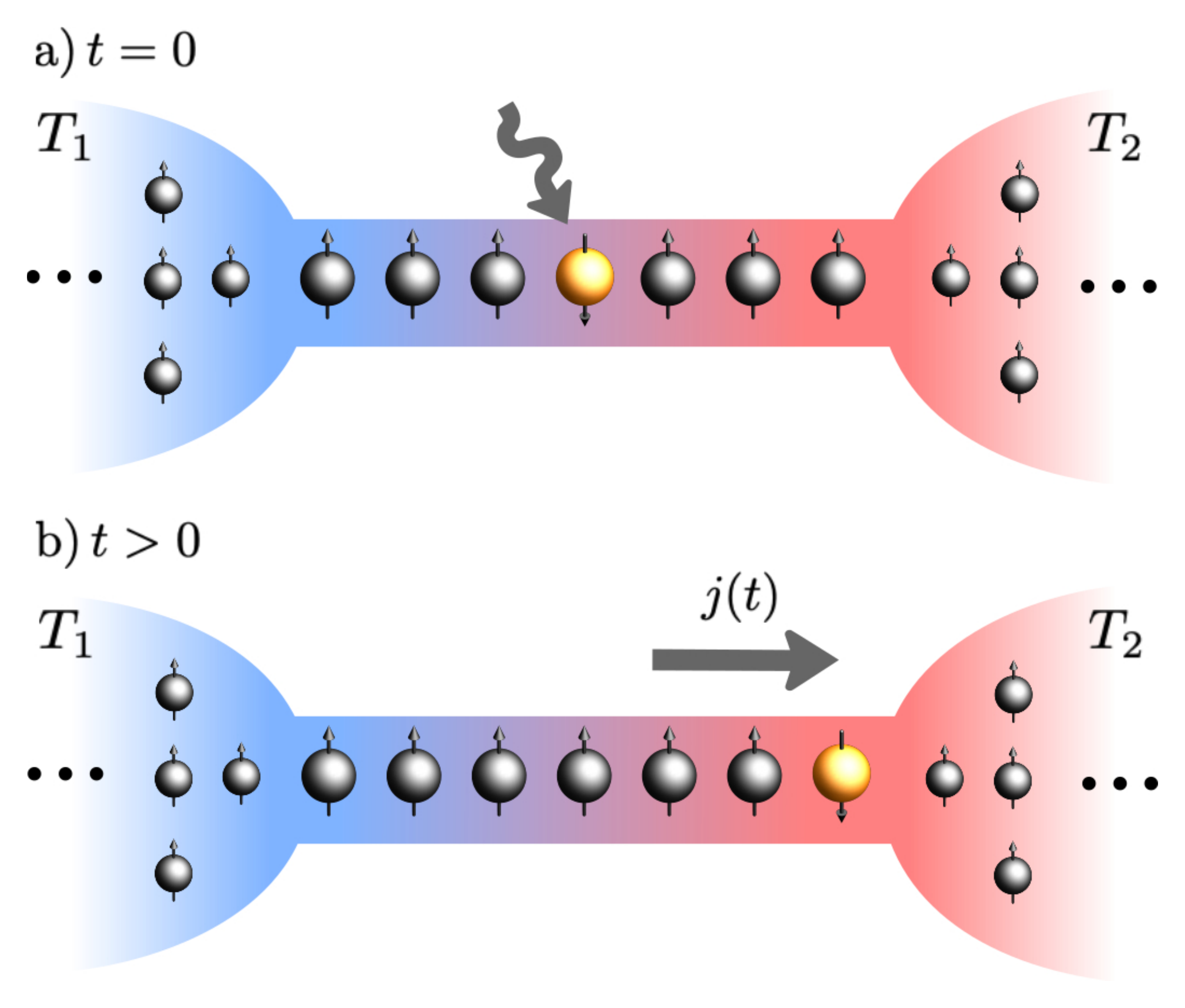}
\caption{Sketch of the system. A small link connects a hot (red, right) reservoir at temperature $T_2$ to a cold (blue, left) reservoir at $T_1$ ($T_2>T_1$). Initially, the system is completely polarized (black `spin-up' particles), and by assumption there is a time-independent temperature gradient across the link. a) At $t=0$ a particle at the center of the system has its internal state changed by a spin-flip pulse. b) Due to the presence of the temperature gradient, one observes a spin current across the system, which is caused by the motion of the impurity from the low- to the high-temperature region.}
\label{fig1}
\end{figure}

\paragraph{System description.} We consider a quasi-one-dimensional quantum wire whose ends are connected to two infinite reservoirs, see Fig.~\ref{fig1}.  The system is spin-polarized at $t<0$ ($t$ for time), i.e., it contains only `spin-up' fermionic particles. The reservoirs are at two different temperatures, $T_1$ and $T_2$, which are kept constant at all times. Moreover, the system is in a steady state at $t\simeq 0$, i.e., there are no mass currents. At $t=0$, there is a spin flip of a particle in the link, which can be implemented in cold-atom experiments using microwave or radio frequency pulses. The spin impurity allows us to introduce a meaningful definition of a spin current, since, with a spin-polarized system, one can study only mass currents. Note that without interactions, the motion of the impurity cannot  be affected by the temperature gradient. Therefore, non-trivial spin dynamics is possible only if particles interact. The goal of this paper is to understand the quench dynamics at $t>0$ assuming that the `spin-up'-`spin-down' interaction is strong. We do not introduce any sudden changes to the trapping geometry of the system and assume that a single spin flip does not influence the density (cf.~\cite{artem2}). In other words, we keep the spatial densities fixed and focus only on the spin dynamics. To validate this assumption, we note that the motion of the density of strongly interacting 1D systems is much faster than time evolution of the spin degrees of freedom -- an effect related to the well-studied phenomenon of spin-charge separation, see, e.g.,~Refs.~\cite{recati,kollath2,Auslaender88,kleine,PhysRevLett.125.190401} for more detail about this effect in cold-atom systems.  This allows us to consider spin dynamics in mesoscopic samples with a fixed density (cf.~\cite{artem3,pu2,Barfknecht2019}).

We remark that the reservoirs in cold-atom experiments are finite, and, thus, we must clarify the meaning of a steady state at $t<0$. Indeed, if one simply prepares a cold-atom system with a temperature mismatch, then particles will first flow from the hot reservoir into the cold reservoir due the difference in chemical potentials, in agreement with the Landauer picture~\cite{Landauer1999}. Later, the particle current might be reversed due to the difference in the particle number between two reservoirs, and, eventually, the system will come to a thermal equilibrium. We note that the timescale of particle transfer can be tuned in these experiments~\cite{Krinner_2017,Esslinger2020}. In particular, this timescale can be made comparable or smaller than the timescale associated with spin transfer in small strongly interacting one-dimensional systems, which is typically $0.1-1$ms. This makes the steady-state assumption adequate.

We consider the link between the reservoirs as a one-dimensional strongly interacting gas whose size and number of particles is fixed. To simulate the effect of the reservoirs whose properties are not affected by the emission and absorption of particles from the link, we will later employ a Lindblad master equation, with spin-flip operators acting at the edges of the system. These play the role of the impurity being exchanged with a particle from a reservoir. We shall also include a `pump' term, which realizes a spin-flip at the center of the chain, in contrast to the operators acting at the edges. These effects could in principle also be modelled through a closed-system approach, in which the entire system (both reservoirs plus the link) are described by a strongly interacting, low-density, one-dimensional system~\cite{Maslov1995,Matveev2004}. We leave, however, the exploration of this approach to future studies. 

\paragraph{Hamiltonian.}
To model the link disconnected from the reservoirs, we adopt the following `fundamental' Hamiltonian 
\begin{equation}\label{ham}
H=\sum_{i=1}^{N_\uparrow} h_{\text{s}}(x_i)+\sum_{j=1}^{N_\downarrow} h_{\text{s}}(y_j) +g\sum_{i,j}\delta(x_i-y_j),
\end{equation}
where $h_{\text{s}}(x)=-\frac{\hbar^2}{2m}\frac{\partial^2}{\partial x^2}+V(x)$ is the single-particle Hamiltonian.  Below, the trapping potential $V(x)$ is a box potential, although our findings can be extended to inhomogeneous potentials that change weakly on the length scale given by the density of the gas. Particles in the system have identical masses $m$, but can be differentiated by some internal degree of freedom, which we label as $\uparrow,\downarrow$ and refer to as the `spin'.  The total number of particles, $N=N_{\uparrow}+N_{\downarrow}$, is time-independent, but the value of $N_{\uparrow}$ ($N_{\downarrow}$) can be changed, e.g., by a spin-flip protocol (see below). In cold-atom experiments, spin can be simulated using hyperfine~\cite{Guan2013Review} or nuclear spin states \cite{fallani}. The interaction in Eq.~(\ref{ham}) is modelled by a delta-function potential. Its strength $g$ is related to the three-dimensional scattering length and to details of the trapping geometry~\cite{olshanii}. For simplicity, we shall use the systems of units in which $\hbar=m=1$.

We assume that particles in the link are strongly interacting, in the sense that the energy scale associated with interactions is much larger than any other energy scale of the problem. This limit is often denoted as $g\to\infty$. It can be simulated with cold atoms at the few- and many-body levels~\cite{weiss2,paredes,jochim1,jochim3}.
For $1/g=0$, the Hamiltonian~(\ref{ham}) can be related to a problem of $N_{\uparrow}+N_{\downarrow}$ spin-polarized fermions~\cite{girardeau}. For strong but finite interactions, it was shown in Refs.~\cite{artem1,deuretz2,artem2} that the solution of the problem can be obtained by diagonalizing the spin-chain Hamiltonian:
\begin{equation}\label{spinchain}
\mathcal{H}_A=E_A-\frac{1}{2g}\sum_{l=1}^{N-1}\alpha_{A;l}(1-\vec{\sigma}_l\cdot \vec{\sigma}_{l+1}),
\end{equation}
where $\vec{\sigma}_l$ is a vector of Pauli matrices. Here, $A$ denotes a particular energy manifold which determines not only the energy at the fermionization limit $E_A$, but also the value of the exchange coefficients $\alpha_{A;l}$ (which furthermore depend on the choice of the underlying trapping potential). In the Methods section we provide details on how the wave function for a particular manifold $A$ is constructed, as well as how to calculate the coefficients $\alpha_{A;l}$.

To date, the focus of theoretical works was mainly on the spin-chain Hamiltonian with $A=1$, which describes the ground state properties of a strongly interacting system~\cite{massignan,xiaoling,xiaoling2,barfk3,marchukov,loft1,artem3,Pan2017,Volosniev2017MassImbalance}. The coefficients $\alpha_{1;l}$ that enter $\mathcal{H}_1$ have been calculated with great precision in systems as large as $N\approx 60$~\cite{conan,deuretz_mdist}, allowing one to study in depth static and dynamic behavior at $T=0$. In the present paper, we extend the discussion of spin-flip dynamics to excited manifolds, which allows us to study finite-temperature physics.

\paragraph{Exchange coefficients.}
As mentioned previously, we assume that $V(x)$ is a box potential, i.e., the potential is zero if $-L/2<x<L/2$, and infinite otherwise. In this case the exchange coefficients are position-independent for the ground state manifold~\cite{Marchukov2016,Pan2017}, i.e., $\alpha_{1;i}=\alpha_{1;j}$. We have checked that $\alpha_{A;i}=\alpha_{A;j}$ also for excited manifolds (within numerical accuracy~$0.01
\%$), which allows us to simplify the notation: $\alpha_{A}\equiv \alpha_{A;i}$. As an additional check, we have calculated coefficients $\alpha_{A;i}$ for a random set of $\{A;i\}$ using the numerical routine CONAN~\cite{conan}. To the best of our knowledge, there is no rigorous proof that $\alpha_{A;i}=\alpha_{A;j}$ for a general value of $A$, but for $A=1$ one can show that the coefficients $\alpha_{1,i}$ do not depend on $i$ by analyzing the Bethe ansatz solution~\cite{Pan2017}. It seems straightforward to extend the line of argument in Ref.~\cite{Pan2017} to excited-state manifolds. 

We now notice that $\alpha_A=\sum_{i=1}^{N-1}\alpha_{A;i}/(N-1)$, which is a trivial identity if $\alpha_{A;i}=\alpha_A$. However, this expression for $\alpha_A$ allows us to integrate over the whole space $x_i\in [-L/2,L/2]$ (see  Eq.~(\ref{geo}) in the Methods section). Once the boundaries of a specific ordering of particles do not play a role, the calculation of $\alpha_A$ is simple: $\alpha_A=2 E_A/ L$. 

The quantity $\alpha_A/g$ is the only parameter in Eq.~(\ref{spinchain}) that can define a non-trivial time scale for the spin dynamics. We use this fact in the Methods section to show that the spin dynamics at low temperatures are determined by the Hamiltonian
\begin{equation}
    h=\frac{\alpha(T)}{2g}\sum_{l=1}^{N-1}\vec{\sigma}_l\cdot \vec{\sigma}_{l+1},
    \label{eq:h_map}
\end{equation}
where $\alpha(T)=2\epsilon(T)/L$, and $\epsilon(T)=\sum_{A}P_A(T)E_A$ is the average energy of a system of $N$ spinless fermions at temperature $T$. Our derivation shows that the exchange coefficient $\alpha(T)$ is a natural extrapolation of the zero-temperature coefficient, $\alpha(0)=2\epsilon(0)/L$~\cite{Pan2017,Volosniev2017MassImbalance}, to finite $T$. Note that $\alpha(T)$ is an increasing function of $T$, which means that that the spin dynamics becomes faster as we increase the temperature. This is logical -- when we increase $T$, we increase the kinetic energy, which makes an exchange of particles quicker.

We propose to use the Hamiltonian~(\ref{eq:h_map}) for the analysis of the time dynamics in a strongly interacting system at finite temperatures. The advantage of this Hamiltonian [for instance, with respect to Eq.~(\ref{eq:decomposition})] is that it can be analyzed using well-developed approaches to the spin-chain Hamiltonians, see, e.g.,~\cite{Karabach1997}. In particular, the spin-flip dynamics can be studied using the spectrum of magnons (see Methods section for details).

Let us take a moment to discuss the dependence of the coefficients $\alpha(T)$ on temperature in the thermodynamic limit. For small temperatures, we expand the energy $\epsilon(T)$ using the Sommerfeld expansion~\cite{ashcroft1976solid}, see also~\cite{cetina1977}, 
\begin{equation}
    \frac{\epsilon(T)}{L}=\frac{\pi^2\rho^3}{6}+\frac{1}{6\rho} (k_{\text{B}} T)^2,
\end{equation}
where $\rho=N/L$ is the density of the gas and $k_{\text{B}}$ is the Boltzmann constant. The corresponding expansion for $\alpha$ reads as
\begin{equation}\label{analyticT}
    \alpha(T)\approx \alpha_1\left[1+\frac{1}{\rho^4 \pi^2} (k_{\text{B}} T)^2\right].
\end{equation}
where $\alpha_1=\alpha(T=0)$ (that is, the exchange coefficient calculated using solely the ground state manifold). The expression above shows a $T^2$-dependence of the coefficients $\alpha(T)$ in this limit. For high temperatures, the equipartition theorem requires $\epsilon(T)$ be proportional to $T$. Note that we do not consider this high-$T$ limit in the paper, since the mapping onto the effective Hamiltonian~(\ref{eq:h_map}) fails when the values of $\alpha(T)$ are large, see Methods section. 

\begin{figure}
\centering
\includegraphics[width=0.5\textwidth]{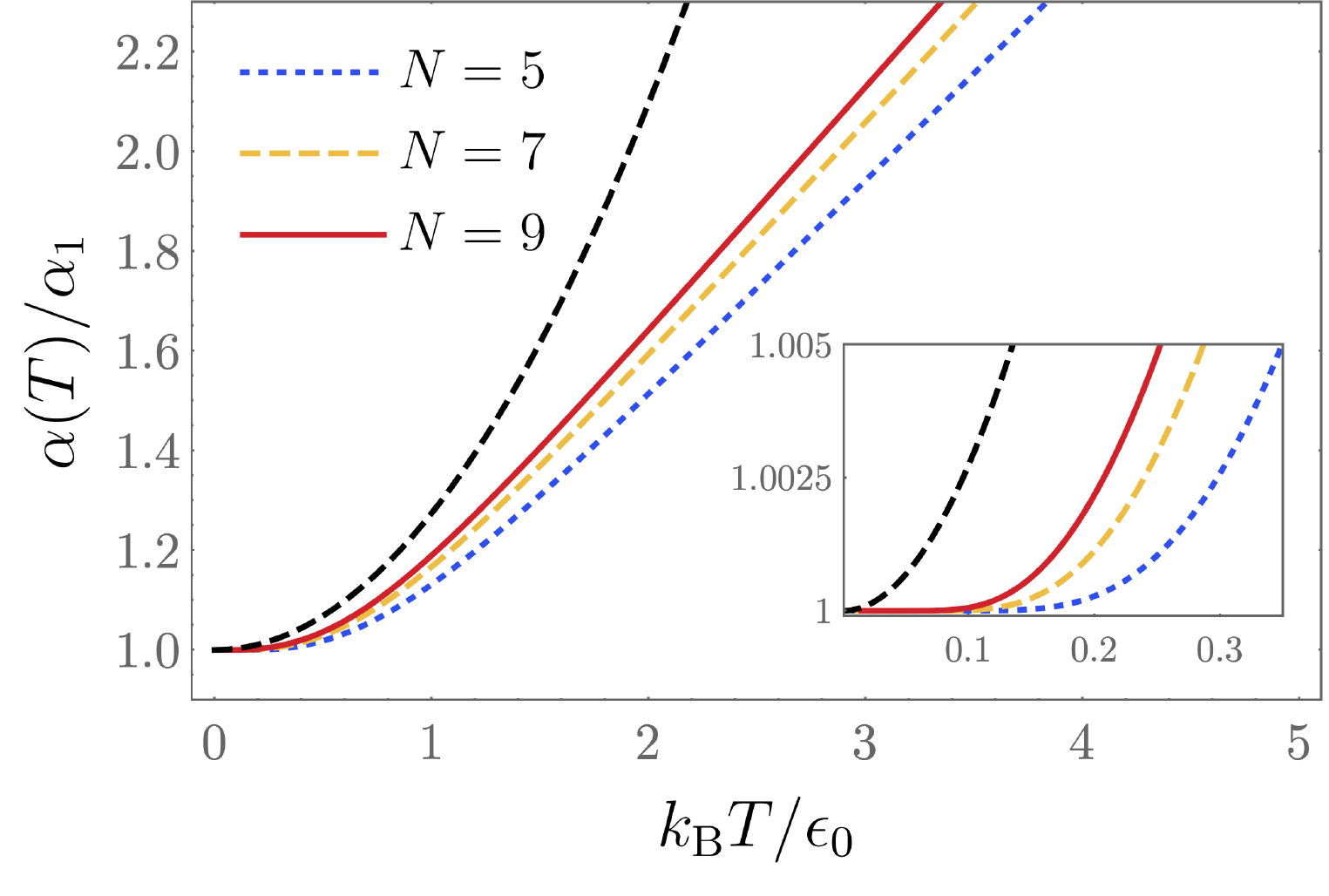}
\caption{Temperature dependence of the exchange coefficients. Exchange coefficients $\alpha(T)$ for different numbers of particles $N=5,7$ and $9$ (dotted blue, dashed yellow and solid red curves, respectively). Here, $\alpha_1$ denotes the values of the coefficients at $T=0$. The black dashed curve shows the prediction of Eq.~\eqref{analyticT}. The inset zooms in the region~$k_{\text{B}} T/\epsilon_0 \ll 1$, where $k_{\text{B}}$ is the Boltzmann constant and $\epsilon_0$ is the energy per particle in the thermodynamic limit. All presented quantities are dimensionless.}
\label{fig2}
\end{figure}

Finally, we discuss $\alpha(T)$ for a finite number of particles, since we are mainly interested in experimentally relevant small links between the reservoirs. In Fig.~\ref{fig2}, we show the behavior of $\alpha(T)$ for $N=5,7$ and $9$. We also plot the low-temperature result of Eq.~(\ref{analyticT}). For convenience, we normalize $k_{\text{B}}T$ by the energy per particle in the thermodynamic limit (with density $\rho=1$), that is $\epsilon_0=\pi^2/6$. 
All in all, the results for finite $N$ qualitatively agree with the predictions of Eq.~\eqref{analyticT} for $k_{\text{B}} T/\epsilon_0 \ll 1$; in other words, finite-size effects lead only to a quantitative change. To ensure the convergence of the curves in Fig.~\ref{fig2}, we use $10^6$ energy manifolds. Smaller values of the number of manifolds (for example, $\simeq 10^4$) do not lead to accurate results for $\alpha(T)$ for the considered parameters. For the time dynamics, this implies that a simultaneous consideration of $10^6$ energy manifolds is needed. The effective Hamiltonian $h$ provides a convenient way to incorporate these many manifolds. It is worthwhile noting that $\alpha(T)/\alpha_1\simeq 1$ for $k_{\text{B}} T/\epsilon_0\lesssim 2$; therefore, a necessary condition for the validity of the mapping of $\mathcal{H}$ onto $h$ can be satisfied in an experimentally accessible window of temperatures.  

In the following, we will use Eq.~\eqref{eq:h_map} along with the local density approximation to study dynamical properties of an impurity in the presence of a temperature gradient. Note that we assume that the local density is constant across the system even with the temperature gradient. This assumption is crucial, and should hold true for experimental set-ups, which attempt to measure the discussed spin dynamics.  The Hamiltonian $h$ can be used to describe also spin dynamics that follow more than a single spin flip (arbitrary number of magnons). However, it should not be used to study static properties. In the derivation of Eq.~\eqref{eq:h_map}, we explicitly rely on the time-dependent nature of the problem at hand. Therefore, we do not expect that Eq.~\eqref{eq:h_map} can describe static properties accurately.

\paragraph{Dynamics in the presence of a temperature gradient.} Once we have established the mapping given in Eq.~\eqref{eq:h_map}, we can use it to investigate the quench dynamics. Quench dynamics are considered to understand transport properties of the many-body Heisenberg model and related Hamiltonians, especially the transition from ballistic to diffusive regimes, see, e.g.,~\cite{PhysRevB.79.214409,PhysRevLett.106.220601,Mendoza_Arenas_2013,Jepsen2020}. Here we study quench dynamics to investigate the effect of a temperature gradient on the motion of an impurity. We stress that, within our formalism, the simplest case of a constant temperature field across the system leads only to faster dynamics in comparison to the zero-temperature limit, and does not introduce any additional effects. The temperature gradient is essential for the findings discussed in this section. 

In this section, we illustrate the dynamics in finite systems (see the Methods section for a brief discussion of the current in the thermodynamic limit). To that end, we study time evolution that follows a spin flip in a small link with $N=7$. First, we consider a closed system where the link is decoupled from the reservoirs, and then an open system where the reservoirs are modelled using the Lindblad master equation. In all calculations, we assume dimensionless time and temperature units by writing $J_0t$ and $k_{\text{B}} T/\epsilon_0$ respectively, where $J_0=\alpha_1/g$. The corresponding time scale for cold-atom experiments can be tuned by changing the density of particles and the value of $g$. To interpret our results in this section, one could use values $0.1-1$ms, which are typical for cold alkali atoms, e.g., $^6$Li.

\paragraph{Closed system.} We start our study of the spin-fip dynamics by assuming
a closed system where the temperature gradient is introduced in the form of a simple step function. We consider the effective model~\eqref{eq:h_map} with the exchange coefficients 
\[\alpha_i= \begin{cases} 
      \alpha(T_1), & i < (N+1)/2 \\
      \alpha(T_2), & i \geq (N+1)/2 \\
   \end{cases},
\]
where $N=7$, $1 \leq i \leq N-1$, and we fix $k_{\text{B}} T_1=0$ and $k_{\text{B}} T_2=2 \epsilon_0$. For convenience, we define the temperature difference as $\Delta T =(T_2-T_1)$. For this particular system size, we have $\epsilon_0/\epsilon_{\text{F}}=1/3$, where $\epsilon_{\text{F}}$ is the Fermi energy. 

\begin{figure}[H]
\centering
\includegraphics[width=0.55\textwidth]{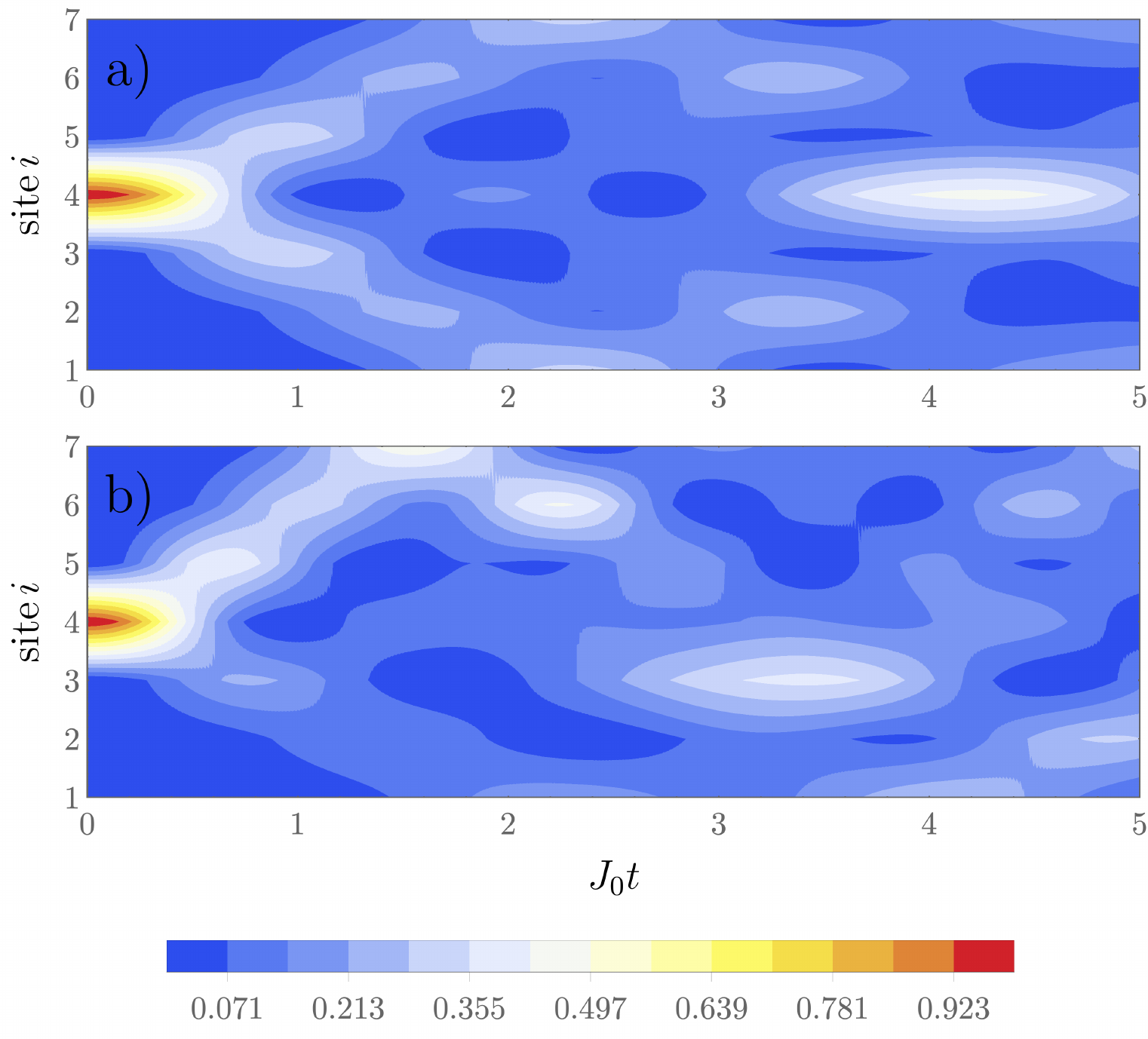}
\caption{Time evolution of the impurity. The contour plot presents the average probability for the impurity to be found at the $i$th-site, $\langle S_{i}^{\downarrow}(t)\rangle$, as a function of time, $t$. The data are for an $N=7$ system with an impurity initialized at the center with a)~$k_{\text{B}} \Delta T=0$ and b)~$k_{\text{B}} \Delta T=2 \epsilon_0$. Here, $\epsilon_0$ is the energy per particle in the thermodynamic limit, $k_{\text{B}}$ is the Boltzmann constant, $J_0=\alpha_1/g$, where $\alpha_1$ is the value of the exchange coefficients at $T=0$ and $g$ is the interaction strength. In panel a), the impurity probability evolves symmetrically with respect to the edges as the system is homogeneous. In panel b), the presence of the temperature difference across the system leads to a directional motion of the impurity towards the high-temperature edge. Red (blue) colors correspond to higher (lower) probabilities. All presented quantities are dimensionless.}
\label{fig3}
\end{figure}

For the sake of discussion, we take as the initial state $\vert \psi_0\rangle=\vert\uparrow\uparrow\uparrow\downarrow\uparrow\uparrow\uparrow\rangle$, and then consider time evolution of this state under the effect of the temperature difference, i.e., we solve the Schr{\"o}dinger equation $i\hbar\psi'=h\psi$ with $\psi(t=0)=\psi_0$. We start by calculating the average probability for the impurity to be found at a given site,
\begin{equation}
    \langle S_{i}^{\downarrow}(t)\rangle=\langle \psi(t)\vert S_{i}^{\downarrow}\vert \psi(t)\rangle,
\end{equation}
where $S_{i}^{\downarrow}=(\mathbb{1}-\sigma^z_i)/2$. 
Figure~\ref{fig3} illustrates the expectation value of this observable for $k_{\text{B}} \Delta T=2 \,\epsilon_0$, and the zero-temperature case. The temperature difference leads to a higher probability for the motion of the impurity towards the edge with a higher temperature, in agreement with our discussion for the thermodynamic case in the Methods section. 

\begin{figure}
\centering
\includegraphics[width=0.5\textwidth]{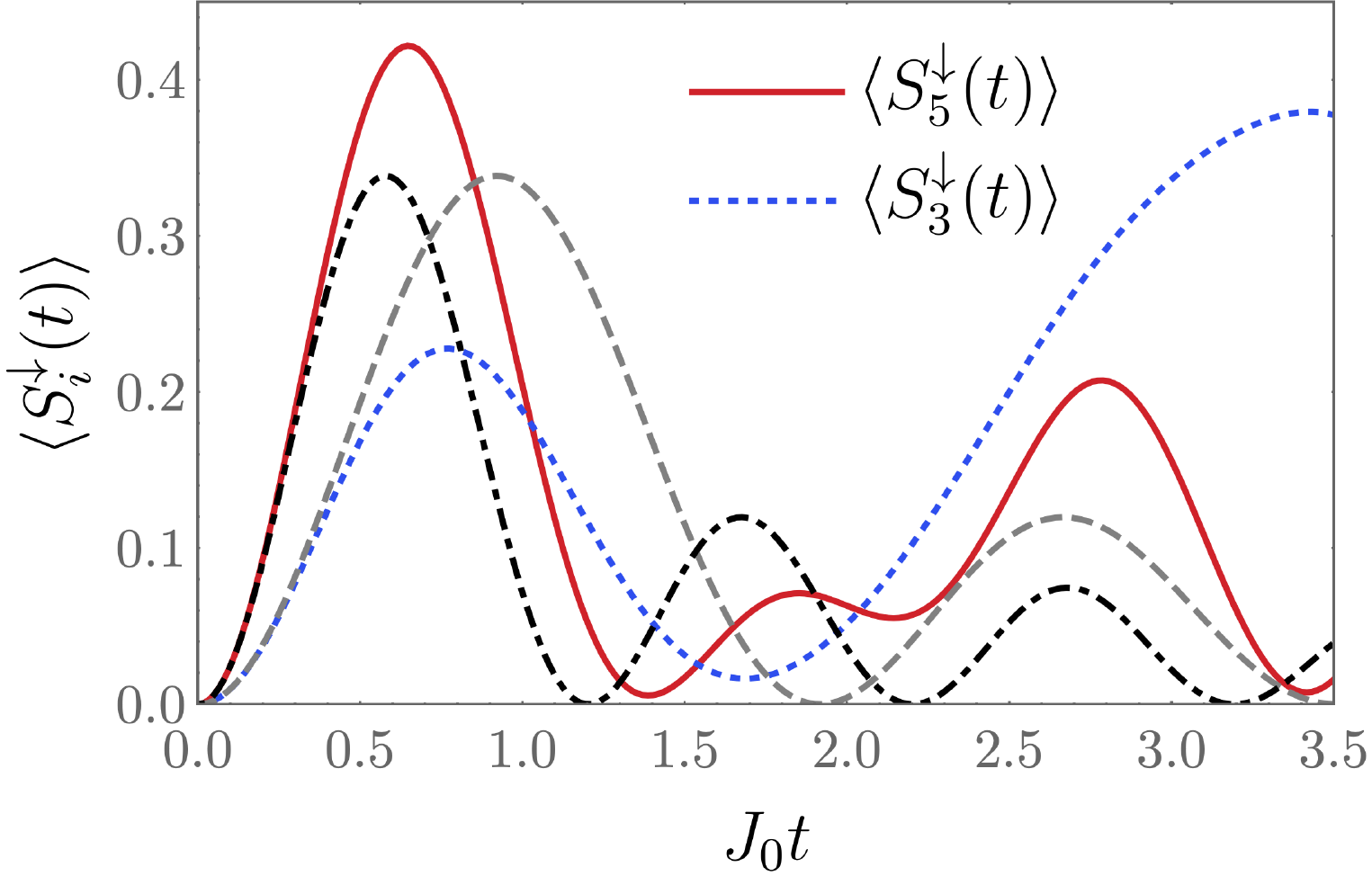}
\caption{Time evolution of the impurity at the sites adjacent to the center of a chain with $N=7$. Here, $k_{\text{B}} \Delta T=2 \epsilon_0$, where $k_{\text{B}}$ is the Boltzmann constant and $\epsilon_0$ is the energy per particle in the thermodynamic limit. The red solid (blue dotted) curve shows the result obtained with Eq.~\eqref{spinchain} for $\langle S_5^\downarrow(t)\rangle$ ($\langle S_3^\downarrow(t)\rangle$). The remaining curves show the predictions of Eq.~\eqref{bessel} with $\alpha=\alpha(T_1)$ (gray, dashed) and $\alpha=\alpha(T_2)$ (black, dot-dashed) at $i=5$. Note that Eq.~\eqref{bessel} implies identical probabilities for $i=3$ and $i=5$. On the horizontal axis, $J_0=\alpha_1/g$, where $\alpha_1$ is the value of the exchange coefficients at $T=0$, and $g$ is the interaction strength. All presented quantities are dimensionless.}
\label{fig4}
\end{figure}

At $\Delta T=0$, the dynamics  of $ \langle S_{i}^{\downarrow}(t)\rangle$ can be found analytically, e.g., by using the Bethe ansatz. The probability of finding the impurity at a particular site for a large system reads as~\cite{Konno2005}
\begin{equation}\label{bessel}
   \langle S_{i-i_0}^{\downarrow}(t)\rangle=\left[\mathcal{J}_{|i-i_0|}\left(J_0 t\right)\right]^2,
\end{equation}
where $i$ denotes the lattice site, $\mathcal{J}$ is the Bessel function of the first kind, and $i_0$ determines the initial position of the impurity. This result is expected to describe a finite system for short times, see also the experiment of Ref.~\cite{fukuhara}. In Fig.~\ref{fig4}, we compare our result for $ \langle S_{i}^{\downarrow}(t)\rangle$ to the predictions of Eq. \eqref{bessel}.  For this comparison, we choose the sites adjacent to the center of the chain $i=\frac{N+1}{2}\pm 1$. 

Figure~\ref{fig4} shows that the temperature gradient introduces an asymmetry in the motion of the impurity. The impurity moves towards the hot side, since large exchange coefficients lead to faster spin dynamics in comparison to the cold side. Figure~\ref{fig4} also compares exact results to those of Eq.~\eqref{bessel} obtained with the coefficients $\alpha(T_1)$ and $\alpha(T_2)$. We conclude that Eq.~\eqref{bessel} can be used to study initial dynamics also for $\Delta T\neq 0$. At later times the inhomogeneity and finite-size effects start to play an important role, and Eq.~\eqref{bessel} fails.

\paragraph{Open system.} We now consider a link in contact with two reservoirs, as depicted in Fig.~\ref{fig1}. To describe the time dynamics of spins in the link, we work with the master equation that describes time evolution of the spin density matrix, $\rho_s(t)$,
\begin{eqnarray}\label{master}
    \frac{\partial \rho_s(t)}{\partial t}=-\frac{i}{\hbar}\left[h,\rho_s(t)\right] +\frac{\gamma}{2} \sum_{i=1,N}\left(2S_{i}^{-}\rho_s(t)S_{i}^{+}-\{\rho_s(t),S_{i}^{+}S_{i}^{-}\}\right),
\end{eqnarray}
where $\left[\cdots\right]$ and $\{\cdots\}$ denote the commutator and anti-commutator, respectively, and the jump operators are given by $S^{\pm}=\left(\sigma^x \pm i \sigma^ y\right)/2$. The coupling to the reservoir is quantified by $\gamma$, which we shall always present in units of $J_0$.  The parameter $\gamma$ describes the rate at which spins up are flipped to spins down (we do not introduce any $\vert \uparrow\rangle \to \vert \downarrow\rangle$ processes at the edges of the link because we assume that the reservoirs are spin polarized). Notice that the coupling to the reservoir occurs only at the edges of the spin chain, and does not depend on temperature (i.e., it is identical at the left and right edges), since we are mainly interested in the dynamics in the link. 

We write the density matrix at $t=0$ as $\rho_s(0)=\vert \psi_0 \rangle \langle \psi_0 \vert$, where $\vert \psi_0 \rangle$ is the initial state considered in the previous subsection, that is $\vert \psi_0\rangle=\vert\uparrow\uparrow\uparrow\downarrow\uparrow\uparrow\uparrow\rangle$. Instead of a step function, now, we consider a linear temperature gradient across the system, which within the local density approximation leads to a set of continuously increasing coefficients $\alpha_i$. The parameter $\Delta T=T_2-T_1$ specifies the difference of the temperatures at the edges of the system. As before, we fix $T_1=0$.  Our main focus is on the total spin current
\begin{equation}\label{current}
    j(t)=\sum_i^{N-1}\langle \psi(t) \vert \left(\sigma_y^i\sigma_x^{i+1}-\sigma_x^i\sigma_y^{i+1}\right)\vert \psi(t)\rangle
\end{equation}
which expresses the net spin motion in the system, and the total magnetization
\begin{equation}\label{mag}
    m(t)=\frac{1}{2}\sum_i^{N}\langle \psi(t) \vert \sigma^z_i\vert \psi(t)\rangle 
\end{equation}
that registers the effect of the loss terms contained in Eq.~\eqref{master}.

\begin{figure}[t]
\centering
\includegraphics[width=1.\textwidth]{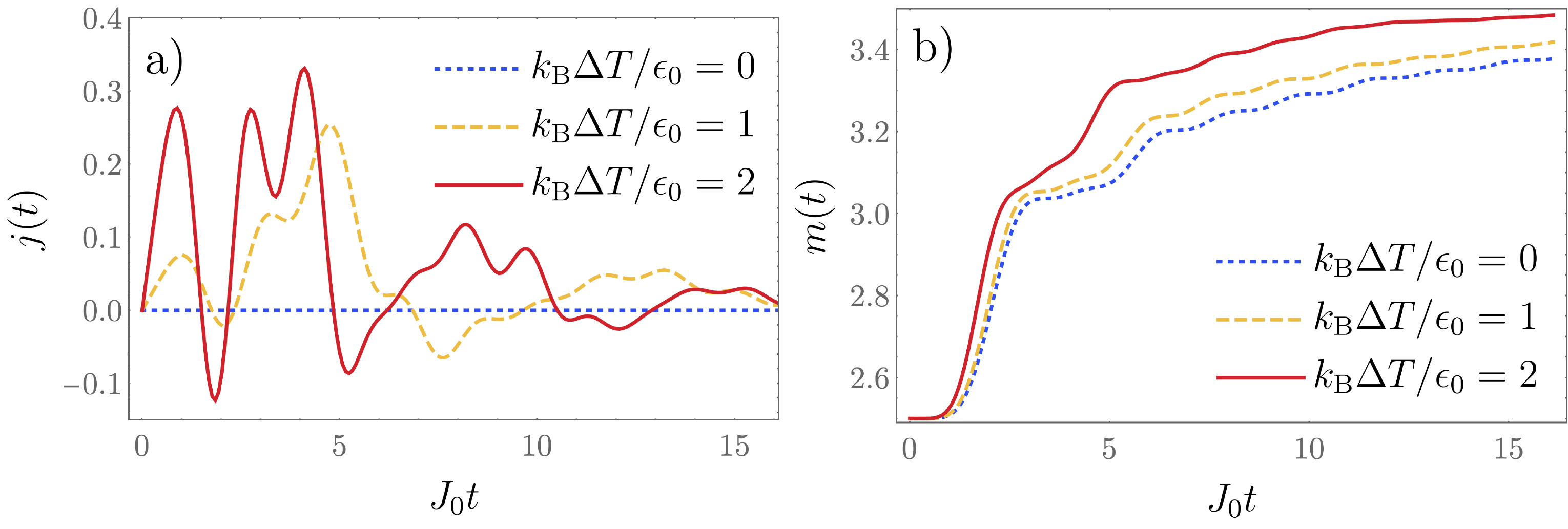}
\caption{Dynamics of spin currents and magnetization. Time evolution, according to Eq.~\eqref{master}, of a) the total spin current and b) the total magnetization with $\gamma/J_0=3.102$ where $J_0=\alpha_1/g$ ($\alpha_1$ is the value of the exchange coefficients at $T=0$ and $g$ is the interaction strength). In both panels $k_{\text{B}}$ is the Boltzmann constant and $\epsilon_0$ is the energy per particle in the thermodynamic limit. Curves show results for temperature gradients indicated in the legends. Our choice of the parameter $\gamma$ (the rate at which spins are flipped at the edges) is arbitrary. It does not change the overall dynamics, but only the time scale for reaching equilibrium. All presented quantities are dimensionless.}
\label{fig5}
\end{figure}

Figure~\ref{fig5} shows time evolution of $j(t)$ and $m(t)$. The spin current occurs in the presence of a finite temperature gradient, see panel~a).  The amplitude of this current is controlled by $\Delta T$. For $t\rightarrow\infty$ the current vanishes due to the effects of the losses at the edges. This can also be detected through the total magnetization: we find that $m(t\to\infty)\rightarrow N/2$, which indicates that the impurity is completely lost to the reservoirs, and the link becomes fully polarized. 

Next, we consider a scenario where the initial state is fully polarized: $\vert\psi(t=0)\rangle=\vert \uparrow\uparrow\uparrow\uparrow\uparrow\uparrow\uparrow\rangle$. The dynamics is initiated by adding the spin-flip term to Eq.~\eqref{master}:
\begin{equation}\label{pump}
    P(t)=\frac{\gamma_{\text{I}}}{2}\left(2S_{\frac{N+1}{2}}^{+}\rho_s(t)S_{\frac{N+1}{2}}^{-}-\{\rho_s(t),S_{\frac{N+1}{2}}^{-}S_{\frac{N+1}{2}}^{+}\}\right),
\end{equation}
which acts only at the center of the chain and can be interpreted as a constant spin `pump' that introduces spin-down spins in the system at a rate given by $\gamma_{\text{I}}$ (in units of $J_0$).

\begin{figure}
\centering
\includegraphics[width=1.\textwidth]{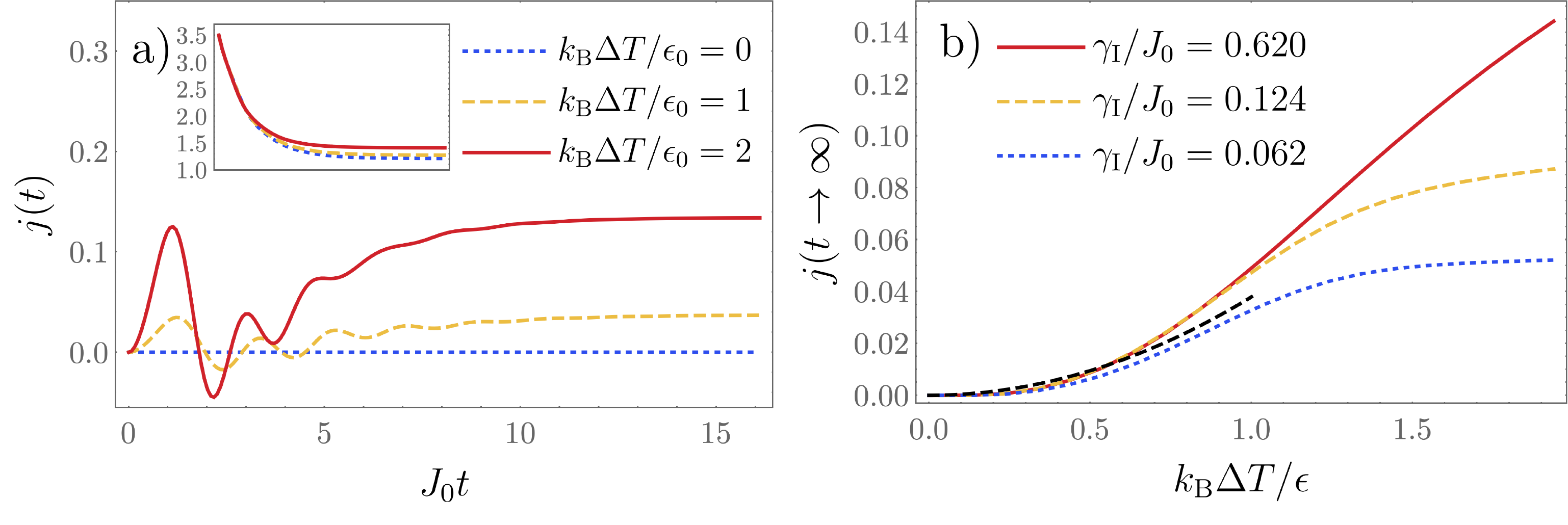}
\caption{Spin currents under the action of a spin pump. a) Time evolution according to Eqs.~\eqref{master} and~(\ref{pump}) of the total spin current for different values of the temperature gradient with $\gamma/J_0=3.102$ and $\gamma_{\text{I}}/J_0=0.620$. Here, $\gamma$ is the loss rate of down spins at the edges (as in the previous plot) and $\gamma_{\text{I}}$ is the rate at which down spins are pumped into the system at the center ($k_{\text{B}}$ is the Boltzmann constant, $\epsilon_0$ is the energy per particle in the thermodynamic limit, $J_0=\alpha_1/g$, where $\alpha_1$ is the value of the exchange coefficients at $T=0$ and $g$ is the interaction strength). The inset shows time evolution of the total magnetization for the same set of parameters. b) The amplitude of the steady-state current as a function of the temperature gradient for different choices of $\gamma_{\text{I}}$. The black dashed curve shows a quadratic fit for $k_{\text{B}}\Delta T/\epsilon_0<1$. All presented quantities are dimensionless.}
\label{fig6}
\end{figure}

Figure~\ref{fig6}~a) presents time evolution of the total current for different choices of $\Delta T$. As in Fig.~\ref{fig5}~a), no current is generated if $\Delta T=0$. For a finite gradient, however, we observe a transient regime for small $t$, evolving towards a steady state at longer times. In the steady state, the losses at the edges match the spin flips in the center. In the inset, we also show the behavior of the magnetization, which, contrarily to the previous case, now drops from the fully polarized value to a constant determined by the parameters of the master equation. It is worthwhile analyzing the behavior of the steady-state current (i.e., $j(t\to\infty)$) for different temperature gradients: we observe an increase in this quantity with $\Delta T$ in all cases, see Fig.~\ref{fig6}~b). For $T\to 0$, this increase is quadratic in temperature, which is in agreement with our previous remark regarding the low-$T$ limit.  We find a noticeable sensitivity of the values of the steady-state current on the parameters $\gamma_{\text{I}}$ and $\gamma$. For example, for very large values of $\gamma_{\text{I}}$, the system becomes saturated with spin-down particles, which reduces the total current. Our conclusion however always holds: a temperature gradient leads to an overall spin current for all considered parameters and protocols.

\section*{Conclusions}\label{concs}

We have studied the dynamics of spin impurities placed in a small link between two reservoirs of different temperatures. Here, the link is described by a one-dimensional system of strongly interacting cold atoms. The reservoirs are simulated using an open quantum system approach. We argue that the dynamics of the system can be obtained by considering a spin chain whose exchange coefficients depend on temperature. Our argument is based upon the Bose-Fermi correspondence and the local density approximation.  

Having established the effective spin-chain Hamiltonian, we consider the motion of a single spin impurity initialized at the center of the link. We observe that the motion of the impurity is highly influenced by the temperature gradient. The impurity moves towards the highest-temperature reservoir, leading to a spin current in the system. Next, we consider a spin pump at the center of the system. In this case, the system evolves towards a steady-state regime with a non-vanishing spin current whose magnitude depends on the temperature gradient. The formalism presented here can be applied to different atomic models, which can be mapped onto spin-chain Hamiltonians. For instance, the same formalism can be applied to study one-dimensional bosonic systems with strong interactions, which realize a XXZ Hamiltonian~\cite{artem2,massignan}. The spectrum of magnons in such a system can be modified by manipulating the boson-boson interactions, and, hence, we expect that the dynamics of an impurity can be made different from that presented here.

Our study provides a microscopic description of the coupling between the spin degree of freedom and temperature for strongly interacting one-dimensional systems.
It paves the way for studying spin caloritronics (and related quantum technologies) with quantum simulators, in particular, cold-atom simulators. 

Even though the present paper focuses on a system of cold atoms, our results are general and can be applied to other physical systems described by the fundamental Hamiltonian. For example, quantum wires at low electron densities provide a possible realization of the model introduced in Eq.~(\ref{ham}). Indeed, in a one-dimensional geometry, low electron density implies strong interactions, which can be described using zero-range potentials. GaAs devices are especially interesting in this regard, for review see~\cite{BEENAKKER_1991,Berggren2010}, and references therein. These devices are some of the cleanest and most studied semiconductor systems. They have relatively weak spin-orbit coupling, which is a prerequisite of using Eq.~(\ref{ham}). Spin coherence necessary for our study is present in the system at temperatures that are low in comparison with the exchange energy between neighbouring electrons. To introduce a spin polarization into the system, one can use a magnetic field. The listed properties put GaAs set-ups forward as solid-state systems for observing the physics discussed here.

\section*{Methods}\label{methods}

\paragraph{Mapping a strongly interacting atomic system onto a spin chain.} We assume that the spectrum of spin-polarized fermions is  $\{E_1,E_2,...\}$ ($E_1\leq E_2\leq E_3\leq ...$); the set of the corresponding many-body wave functions is given by $\{\Psi_1,\Psi_2,...\}$. The function $\Psi_A$ yields $(N_{\uparrow}N_{\downarrow})!/N_{\uparrow}!N_{\downarrow}!$ wave functions of $H$ each with the energy $E_A$. These functions can be written as 
\begin{equation}
\phi_{A,i}(x_1,...,y_{N_{\downarrow}})=\sum_{P=1}^{\frac{(N_{\uparrow}N_{\downarrow})!}{N_{\uparrow}!N_{\downarrow}!}}a_P\Psi_A(x_1,...,y_{N_{\downarrow}}) \mathbf{1}_{G_P},  
\label{eq:phiAi}
\end{equation}
where $\mathbf{1}_{G_P}$ is an indicator function,  $G_P$ determines a specific ordering of particles, e.g., $G_1=x_1<x_2<x_3<...<y_{N_{\downarrow}}$. The coefficients $a_P$ can be calculated by considering $1/g\neq 0$, where
the degeneracy of states is lifted. 

For strong (but finite) interactions, the energies and coefficients $a_P$ are obtained either using perturbation theory around infinite-interaction point~\cite{artem1,Gharashi2015Impurity} or, equivalently, by diagonalizing the Heisenberg Hamiltonian for a given magnetization $(N_{\uparrow}-N_{\downarrow})/2$~\cite{deuretz2,artem2}

\begin{equation}
\mathcal{H}_A=E_A-\frac{1}{2g}\sum_{l=1}^{N-1}\alpha_{A;l}(1-\vec{\sigma}_l\cdot \vec{\sigma}_{l+1}),
\end{equation}
The exchange coefficients $\alpha_{A;l}$ are determined solely by the trapping geometry as
\begin{equation}\label{geo}
\alpha_{A;l}=\int_{x_1<...<x_N-1} dx_1\,...\,dx_{N-1}\Big|\frac{\partial \Psi_A(x_1,...,x_N)}{\partial x_N}\Big|^2_{x_N=x_l}.
\end{equation}
In the limit $g\to \infty$, the eigenstates of the Hamiltonian~(\ref{ham}) are given by the set $\{\phi_{A,i}\}$; the corresponding energies are $\epsilon_{A,i}$ The parameter $A$ determines the energy manifold (note that $\epsilon_{A,i}\simeq E_{A}$ for all values of $i$), and $i$ determines the position of the state within this manifold. We introduce the following notation to express this fact 
\begin{equation}
    H\simeq \sum_{A}\mathcal{H}_A,
    \label{eq:decomposition}
\end{equation}
which means that eigenstates in $H$ can be obtained using eigenstates of $\mathcal{H}_A$, which are presented in Eq.~(\ref{eq:phiAi}). This is true as long as the gap between $E_A$ and $E_{A\pm 1}$ is large enough, i.e., the state $A$ is not coupled by particle-particle interaction to the states with $A\pm 1$. Such a decoupling is essential for our discussion. It naturally occurs in few-body systems with large values of $g$ even for highly excited states, i.e., for $A\gg 1$.    

\paragraph{Dynamics at finite temperature.} We study here the quench dynamics that follow a spin-flip in the link (assuming that the link is disconnected from the reservoirs) without a temperature gradient. We consider the following spin-flip protocol at $T\neq 0$: At $t<0$, there are $N$ spin-polarized fermions at temperature $T$, which are described by the density matrix 
\begin{equation}
    \rho=\sum P_A(T)|\Psi_A\rangle \langle \Psi_A|,
    \label{eq:initial}
\end{equation}
where $|\Psi_A\rangle$ is a many-body state whose spatial representation is $\Psi_A$. $P_A(T)=e^{-\beta E_A}/\sum_A e^{-\beta E_A}$, where $\beta=1/(k_{\text{B}} T)$. At $t=0$, a single spin is flipped somewhere in the system. This could be a single-site flip or a quantum superposition involving a few sites. Our goal is to understand the time dynamics at $t>0$. According to Eqs.~(\ref{eq:decomposition}) and~(\ref{eq:initial}), any spin observable $\mathcal{O}(t)$ can be calculated by considering different manifolds separately, i.e.,
\begin{equation*}
    \mathcal{O}(t)=\sum P_A(T)\mathcal{O}_A(t),
\end{equation*}
where $\mathcal{O}_A(t)$ describes the time dynamics of the observable in a given manifold. The unitary map that determines the corresponding time dynamics reads as $e^{-i\mathcal{H}_At}$. The time evolution of any observable can be written as $\mathcal{O}_A(t)=f(\alpha_A t)$, where $f$ is some function that depends on $\mathcal{O}_A(t)$. The observable $\mathcal{O}$ is then given by
\begin{equation}
    \mathcal{O}(t)=\sum_{A}P_A(T)f(\alpha_A t).
\end{equation}
Our focus is on low temperatures, for which only low-energy states are populated, and, hence, $\alpha_A=\alpha_1(1+\delta_A)$, where $\delta_A\ll 1$. Using that $\sum P_A=1$, we rewrite $\mathcal{O}$ as 
\begin{equation}
\mathcal{O}(t)=f\left(t \sum_A P_A(T)\alpha_A\right)+\sum_A O((\delta_A t)^2).
\end{equation} 
This expression allows us to obtain the spin dynamics - up to the terms $O((\delta_A t)^2)$ - governed by $\mathcal{H}$ from the time dynamics resulting from the effective Hamiltonian (see Eq.~(\ref{eq:h_map}))
\begin{equation}
    h=\frac{\alpha(T)}{2g}\sum_{l=1}^{N-1}\vec{\sigma}_l\cdot \vec{\sigma}_{l+1}.
\end{equation}

As mentioned in the main text, the spin-flip dynamics in our XXX model can also be studied using the spectrum of magnons:
\begin{equation}
    \mathcal{E}_M(p,T)= \frac{2\alpha(T)}{g} (1-\cos(p)),
    \label{eq:energy_magnon}
\end{equation}
where $p$ is a quasi-momentum and $p=2\pi n/N$ with integer $n$. In the infrared limit of $N\gg n$, we can write $\mathcal{E}_M(p,T)\simeq \alpha(T)p^2/g$, which allows us to study the dynamics using a free-particle picture with the temperature-dependent effective mass: $m_{\mathrm{eff}}=g/(2\alpha(T))$. This renormalization of the effective mass of a magnon can be observed in cold-atom experiments.
In particular, if we assume that at $t=0$ the wave packet of a spin-impurity has a Gaussian profile, then the probability density at later times is
\begin{equation}
    |\psi(x,t)|^2=e^{-\frac{2x^2}{1+4 t^2/m_{\mathrm{eff}}^2}}\sqrt{\frac{2}{\pi(1+4 t^2/m_{\mathrm{eff}}^2)}}.
\end{equation} 
This formula shows how the effective mass changes the time dynamics, which can be detected in situ~\cite{fukuhara}. By exploring a single-particle picture further, we conclude that the experiment should observe a diffusive behavior [in a sense of an analytic continuation to classic diffusion] of an initial spin flip with the diffusion coefficient that depends on the temperature as: $D=[2m_{\mathrm{eff}}(T)]^{-1}$.  

\paragraph{Dynamics in a temperature gradient: results in the thermodynamic limit.} Here we consider an infinite link ($L\to\infty$) where the temperature gradient is defined by a step function, i.e., $T=T_1$ for $x<0$ and $T=T_2$ for $x>0$. We apply the local density approximation to write $\alpha(T_1)$ [$\alpha(T_2)$] for the exchange coefficients at $x<0$ [$x>0$]. Let us then investigate what happens to the impurity as it is introduced in the center of the system (at $x\simeq 0$). We start by formulating two observations that might suggest different outcomes: on the one hand, higher temperatures lead to faster dynamics, favoring movement of the impurity into the high-temperature region. On the other, the density of states in the high-temperature region is smaller than that in the colder region, indicating the opposite dynamics. Without the temperature gradient, the density of states may be calculated using Eq.~(\ref{eq:energy_magnon}). Assuming a large system, we derive that the density of states for $x>0$ is proportional to $\sqrt{g/(E\alpha(T_2))}$ and for $x<0$ it scales as $\sqrt{g/(E\alpha(T_1))}$. To estimate the relative importance of these observations, we consider the Schr{\"o}dinger equation
\begin{equation}
    -\frac{1}{2m_{\mathrm{eff}}(x)}\frac{\partial^2}{\partial x^2}f=E f,
\end{equation}
which describes the infrared dynamics of the impurity in the thermodynamic limit; $m_{\mathrm{eff}}=g/(2\alpha(T_1))$ for $x<0$ and $m_{\mathrm{eff}}=g/(2\alpha(T_2))$ for $x>0$. The solution we are after reads as 
\[f= \begin{cases} 
      e^{-ik_1 x}, & x < 0 \\
      e^{i k_2 x}, & x >0  \\
   \end{cases},
\]
where $k_1^2\alpha(T_1)=k_2^2\alpha(T_2)=g E$. These expressions constitute a phenomenological description 
of a source of particles with a given energy $E$ at $x\simeq 0$. The flux that corresponds to these solutions is given by $j_P(x<0)=\sqrt{4E\alpha(T_1)/g}$ and $j_P(x>0)=\sqrt{4E\alpha(T_2)/g}$. We see that the particle is more likely to move into the region with high temperature. The ratio of the probability currents reads as $j_P(x>0)/j_P(x<0)=\sqrt{\alpha(T_2)/\alpha(T_1)}$:
\begin{equation}
\frac{j_P(x>0)}{j_P(x<0)}\simeq 1+(T_2-T_1)(T_2+T_1)\frac{k_{\text{B}}^2 m^2}{2\hbar^4\rho^4\pi^2}.
\label{eq:ratio_currents}
\end{equation}
Notice the quadratic dependence of the currents on temperature for $T_1\to 0$. This dependence is typical for the low-temperature spin currents in our model, and is verified in our simulations. 

Finally, we note that in the derivation above we fixed the energy $E$. If we fixed the momentum instead, which is logical if the parts with different temperatures are disconnected, then we would derive that $j_P(x>0)/j_P(x<0)=\alpha(T_2)/\alpha(T_1)$. This modification does not change conclusions of this subsection.

\section*{Data availability}
The datasets generated during and/or analysed during the current study are available from the corresponding author on reasonable request.

\bibliographystyle{naturemag}
\bibliography{biblio}

\section*{Acknowledgements}
The authors acknowledge support from the European QuantERA ERA-NET Cofund in Quantum Technologies (Project QTFLAG Grant Agreement No. 731473) (R.E.B),  CNPq (Conselho Nacional de Desenvolvimento Científico e Tecnológico) Brazil (A.F.), the European Union's Horizon 2020 research and innovation programme under the Marie Sk\l{}odowska-Curie Grant Agreement No. 754411 (A. G. V.), the Independent Research Fund Denmark, the Carlsberg Foundation, and Aarhus University Research Foundation under the Jens Christian Skou fellowship program (N.T.Z).

\section*{Author Contributions}
A.F., A.G.V. and N.T.Z. devised the project, and R.E.B and A.G.V developed the formalism. The calculations were carried out by R.E.B and A.G.V.. The initial draft of the manuscript was written by R.E.B. and A.G.V.. All authors contributed to the revisions that led to the final version.

\section*{Additional Information}
The authors declare no competing interests.

\end{document}